\documentclass[twocolumn,showpacs,preprintnumbers,amsmath,amssymb,prb]{revtex4}

\usepackage{graphicx}
\usepackage{dcolumn}
\usepackage{bm}

\begin{document}

\preprint{APS/123-QED}

\title{Thermodynamic properties of quantum sine-Gordon spin chain system KCuGaF$_6$}

\author{Izumi Umegaki$^1$}
\author{Hidekazu Tanaka$^1$}
  \email{tanaka@lee.phys.titech.ac.jp.}
\author{Toshio Ono$^2$} 
\author{Masaki Oshikawa$^3$}
  \author{Kazumitsu Sakai$^4$}

\affiliation{$^{1}$Department of Physics, Tokyo Institute of Technology, Meguro-ku, Tokyo 152-8551, Japan\\
$^{2}$Department of Physics, Osaka Prefecture University, Sakai, Osaka 599-8531, Japan\\
$^{3}$Institute for Solid State Physics, University of Tokyo, Kashiwa, Chiba 277-8581, Japan\\
$^{4}$Institute of Physics, University of Tokyo, Meguro-ku, Tokyo 153-8902, Japan}

\date{\today}

\begin{abstract}
We investigated the thermodynamic properties of the spin-$\frac{1}{2}$ one-dimensional Heisenberg antiferromagnet KCuGaF$_6$ by measuring the specific heat in magnetic fields. When this compound is subjected to a uniform magnetic field $\bm H$, a transverse staggered magnetic field $\bm h$ is induced in this compound owing to the staggered component of$ the \bm g$ tensor and the Dzyaloshinskii-Moriya interaction with an alternating $\bm D$ vector. Consequently, the quantum sine-Gordon (SG) model is an effective model of this compound in a uniform magnetic field. In three different field directions, we observed a magnetic-field-induced gap, which increases with $H$. We analyzed experimental results using specific heat theory based on quantum SG theory. The thermodynamic property for $H\,{\parallel}\,c$ is very well described in terms of the elementary excitations characteristic of the quantum SG model, while for the other field directions, significant contributions from other excitation modes beyond the framework of the quantum SG model were observed. For $H\,{\parallel}\,b$, a quantum phase transition between gapless and gapped ground states was observed.

\end{abstract}

\maketitle


\section{Introduction}

One-dimensional (1D) quantum systems have attracted considerable attention for a long time because of various phenomena resulting from the strong quantum fluctuations characteristic of the 1D systems. In particular, the $S\,{=}\,1/2$ antiferromagnetic Heisenberg chain (AFHC) has been closely studied theoretically. Using the Bethe ansatz, the dispersion relation of the lowest excitation for the $S\,{=}\,1/2$ uniform AFHC, which is known as the des Cloizeaux and Pearson (dCP) mode, has been calculated exactly\,\cite{dCP,Faddeev}. The two-spinon contribution to magnetic excitations has also been calculated.\cite{Yamada,Muller} The energy of the dCP mode is different from the result obtained by linear spin wave theory\cite{Anderson,Kubo} by a factor of ${\pi}/2$, which was verified by inelastic neutron scattering measurements.\cite{Endoh,Nagler} Accurate calculations\cite{Pytte,Ishimura} and inelastic neutron scattering measurements\cite{Heilmann} showed that low-energy excitations in a magnetic field cannot be described even qualitatively by linear spin wave theory. The calculated dispersion relation demonstrated that gapless excitations occur at incommensurate wave vectors $q\,{=}\,{\pm}\,2{\pi}m(H)\,{\equiv}\,{\pm}\,q_0$ and ${\pi}\,{\pm}\,q_0$ in addition to at $q\,{=}\,0$ and ${\pi}$, where $m(H)$ is the magnetization per site in the unit of $g{\mu}_{\rm B}$. 

By means of neutron inelastic scattering and specific heat measurements, Dender {\it et al.}\cite{Dender} observed an unexpected magnetic-field-induced incommensurate gap in Cu(C$_6$H$_5$COO)$_2$$\cdot$3H$_2$O, abbreviated to copper benzoate, which is known as an $S\,{=}\,1/2$ 1D antiferromagnet with good one-dimensionality.\cite{Date1,Date2,Takeda} This problem was discussed by Oshikawa and Affleck\cite{Oshikawa1,Affleck,Oshikawa3} on the basis of the effective Hamiltonian expressed as  
\begin{eqnarray}
\mathcal{H}=\sum_{i} \left[J\bm S_i\cdot \bm S_{i+1}-g{\mu}_{\rm B}HS_i^z-(-1)^ig{\mu}_{\rm B}hS_i^x\right], 
\label{eq:model}
\end{eqnarray}
where $h$ is the transverse staggered field induced by the external field $H$. The staggered field on the $i$th site $\bm{h}_i$ originates from the staggered component of the $\bm g$ tensor $(-1)^i{\bm g}_{\rm s}$ and the Dzyaloshinskii-Moriya (DM) interaction\cite{Moriya} with the alternating $\bm D_i$ vector $(-1)^i{\bm D}$, and is expressed as 
\begin{eqnarray}
\bm{h}_i\simeq(-1)^i\left[\frac{{\bm g}_{\rm s}}{g_{\rm u}}\bm{H}+\frac{\bm{H}}{2J}\,{\times}\,{\bm D}\right],
\label{eq:h_st}
\end{eqnarray}
where $g_{\rm u}$ is the uniform $g$ factor.\cite{Affleck} Because the ${\bm g}_{\rm s}$ tensor and $\bm D$ vector are anisotropic, the proportionality coefficient $c_{\rm s}\,{=}\,h/H$ depends on the field direction. 

\begin{figure}[htbp]
\begin{center}
 \includegraphics[scale =0.5]{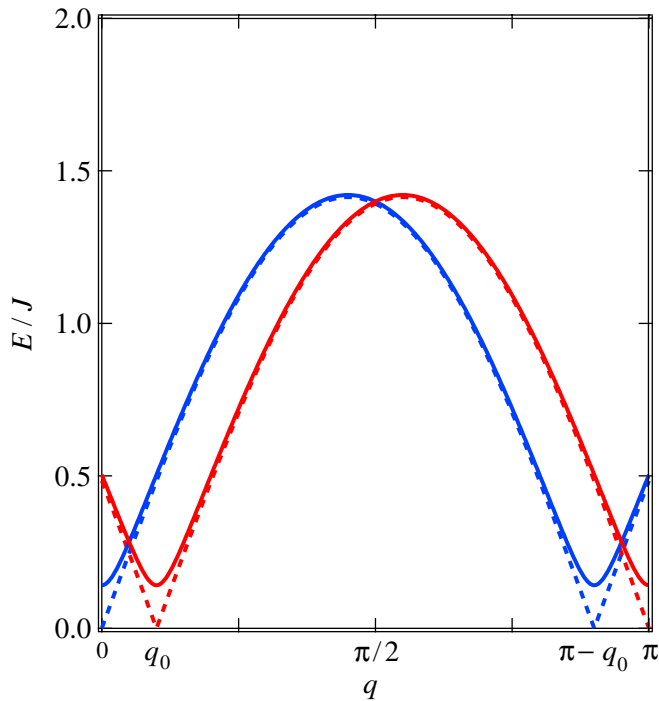}
\end{center}
\caption{Schematic view of the lowest-energy excitations of model (\ref{eq:model}) in a nonzero magnetic field for $\bm h\,{=}\,0$ (dashed lines) and $\bm h\,{\neq}\,0$ (solid lines). Gapless excitations at $q\,{=}\,0$ and at incommensurate waves $q\,{=}\,{\pm}\,q_0$ for $\bm h\,{=}\,0$ have finite gaps for $\bm h\,{\neq}\,0$.} 
\label{fig:excitation_nonzero_mag}
\end{figure}

Using the field-theoretical approach, Oshikawa and Affleck demonstrated that the model given by eq.\,(\ref{eq:model}) can be mapped onto the quantum sine-Gordon (SG) model with Lagrangian density 
\begin{eqnarray}
\mathcal{L}=\frac{1}{2}\left[\left(\frac{{\partial}{\phi}}{{\partial}{t}}\right)^2\,{-}\,{(vJ)}^2\left(\frac{{\partial}{\phi}}{{\partial}{x}}\right)^2\right]+hC\cos (2{\pi}R{\tilde \phi}),
\label{eq:Lag}
\end{eqnarray}
where $\phi$ is the canonical Bose field, $\tilde \phi$ is the dual field, $R$ is the compactification radius, $v$ is the dimensionless spin velocity and $C$ is a coupling constant. 
The dual field $\tilde \phi$ corresponds to the angle between the transverse component of the spin and the reference direction in a plane perpendicular to the external magnetic field. Owing to the nonlinear term in eq.\,(\ref{eq:Lag}), all the excitations are gapped as illustrated in Fig.\,\ref{fig:excitation_nonzero_mag}.

The elementary excitations characteristic of the quantum SG model are the soliton, antisoliton and breather excitations. The soliton excitation corresponds to the excitation at $q\,{=}\,{\pm}\,q_0$ and ${\pi}\,{\pm}\,q_0$, and is classically described as the local rotation of the spin in a plane perpendicular to the magnetic field, and the antisoliton excitation corresponds to the inverse rotation of the spin. The breather corresponds to the bound state of the soliton and antisoliton as reducing their energies with $\tilde \phi \simeq 0$. Oshikawa and Affleck\,\cite{Oshikawa1,Affleck,Oshikawa3} showed that the magnetic-field-induced gap corresponding to the soliton energy is proportional to $H^{2/3}$.
The field dependence of the gap is in good agreement with the experimental result observed for copper benzoate.\cite{Dender} On the basis of quantum SG theory,\cite{Oshikawa3} they also provided a good description for the resonance field of electron spin resonance (ESR) obtained by Oshima {\it et al.}\cite{Oshima2}

The elementary excitations in copper benzoate were investigated in detail using high-frequency ESR.\cite{Asano,Nojiri} In addition to copper benzoate, the following substances are known to be described by the effective model given by eq.\,(\ref{eq:model}): Yb$_4$As$_3$,\cite{Helfrich,Oshikawa2,Kohgi} PM$\cdot$Cu(NO$_3$)$_2\cdot$(H$_2$O)$_2$ (PM\,=\,pyrimidine)\cite{Feyerherm,Zvyagin, Zvyagin2, Wolter1,Wolter2} and CuCl$_2$${\cdot}$2((CD$_3$)$_2$SO).\cite{Kenzelmann1,Kenzelmann2,Chen} Because the elementary excitations and thermodynamic properties of quantum SG systems have been of great interest, numerous theoretical investigations have been published\cite{Essler3,Essler2,KIB,Lou,Capraro,Wang,Essler1,Zhao,Lou2,Glocke,Orignac,Kuzmenko}, and the theoretical results have been used to analyze experimental results on the above-mentioned substances. Moreover, new experiments have been proposed on the basis of theoretical results.\cite{Zvyagin2,Karimi,Xi,Sato}

\begin{figure}[htbp]
	\begin{center}
		\includegraphics[scale =0.45]{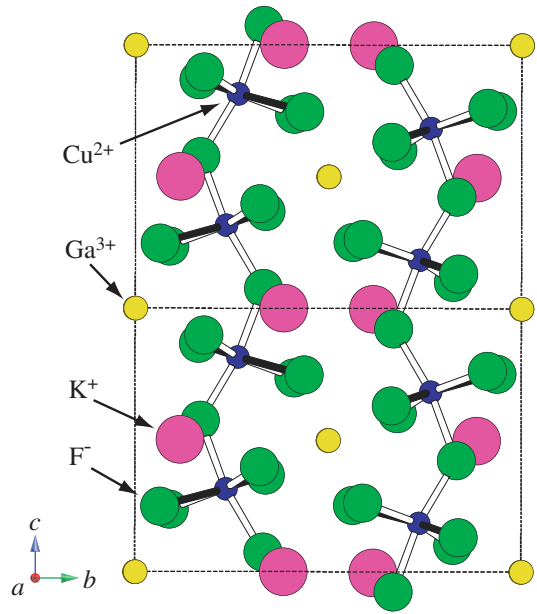}
	\end{center}
	\caption{Crystal structure of KCuGaF$_6$ viewed along the $a$ axis. Dotted lines denote the chemical unit cell. Filled and unfilled bonds respectively denote the elongated and compressed axes of CuF$_6$ octahedra.}
	\label{fig:KCuGaF6}
\end{figure}

KCuGaF$_6$ is an $S\,{=}\,1/2$ AFHC system, which is represented by the effective model given by eq.\,(\ref{eq:model}) in a magnetic field.\cite{Morisaki,Umegaki} KCuGaF$_6$ is composed of corner-sharing CuF$_6$ octahedra running along the $c$ axis, as shown in Fig.\,\ref{fig:KCuGaF6}.\cite{Dahlke} The CuF$_6$ octahedra are elongated perpendicular to the chain direction, which is parallel to the $c$ axis, owing to to the Jahn-Teller effect. Consequently, the hole orbitals of Cu$^{2+}$ ions are linked along the chain direction through the $p$ orbitals of F$^-$ ions, which leads to a strong antiferromagnetic exchange interaction along the $c$ direction of $J/k_{\rm B}\,{=}\,103$ K.\cite{Morisaki,Umegaki} The elongated and compressed principal axes of the octahedra alternate along the chain direction, as shown by the filled and unfilled bonds in Fig.\,\ref{fig:KCuGaF6}, respectively. This gives rise to the staggered component of the $\bm g$ tensor and the DM interaction with the alternating $\bm D$ vector. Because of the $c$ glide plane at ${\pm}\,b/4$, the $ac$ plane component of the $\bm D_i$ vector alternates along the chain direction but the $b$ component does not. Thus, the $\bm D_i$ vector should be expressed as
\begin{eqnarray}
{\bm D}_i=\left[(-1)^iD_x, D_y, (-1)^iD_z\right],
\label{eq:D_i}
\end{eqnarray}  
where the $x$, $y$ and $z$ axes are chosen to be parallel to the $a^*\,({\perp}\,b, c)$, $b$ and $c$ axes, respectively. For these reasons, the staggered transverse magnetic field ${\bm h}_i$ is induced, when this compound is subjected to an external magnetic field $\bm H$. If the $y$ component $D_y$ is negligible, then the $\bm D_i$ vector is expressed as ${\bm D}_i\,{=}\,(-1)^i{\bm D}$. The classical spin arrangements for the staggered and uniform ${\bm D}$ vectors are the canted Neel state and the spiral state, respectively. 

KCuGaF$_6$ differs from other quantum SG substances in its large exchange interaction of $J/k_{\rm B}\,{=}\,103$\,K and its wide range of the proportionality coefficient, $c_{\rm s}\,{=}\,0.03\,{-}\,0.17$.\cite{Morisaki,Umegaki} Thus, KCuGaF$_6$ is expected to be suitable for obtaining a comprehensive understanding of the systems described by model (\ref{eq:model}). In a previous paper,\cite{Umegaki} we observed as many as about ten ESR modes in KCuGaF$_6$ and found that most of the ESR modes can be very well explained by quantum SG field theory.\cite{Affleck,Essler1} The magnetic-field-induced gap in KCuGaF$_6$ was also confirmed by NMR measurement and a gap proportional to $H^{2/3}$ was observed for $H\,{\parallel}\,c$.\cite{Mori}

Specific heat measurement is a powerful tool necessary for obtaining a comprehensive understanding of the excitation nature. In the present paper, we report the results of specific heat measurements on KCuGaF$_6$ in three different field directions. Experimental results are analyzed by performing calculations based on quantum SG field theory. As shown below, the specific heat of KCuGaF$_6$ in a magnetic field can be basically understood in terms of quantum SG field theory. However, we found that additional excitation modes, whose origins are not known, make significant contributions to the specific heat. This paper is organized as follows: in section 2, we summarize the theory of elementary excitations and specific heat ing the quantum SG model. Experimental details are reported in section 3. The experimental results, their analyses and a discussion are given in section 4. Section 5 is devoted to a conclusion. 


\section{Elementary excitations and specific heat in quantum SG model}

\begin{figure}[htbp]
\begin{center}
 \includegraphics[scale =0.65]{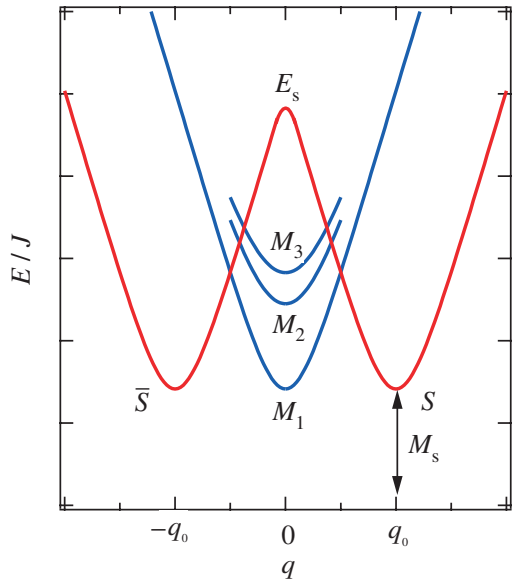}
\end{center}
\caption{Illustration of low-energy excitations of model (\ref{eq:model}) around $q\,{=}\,0$. The soliton, antisoliton, soliton resonance and three breathers are labeled as $S$, $\bar S$, $E_{\rm s}$ and $M_i$ ($i\,{=}$\,1, 2 and 3), respectively.} 
\label{fig:excitations}
\end{figure}

The elementary excitations characteristic of quantum SG model are solitons, antisolitons and their bound states called breathers.\cite{Oshikawa1,Affleck} In field-theoretical language, the excitation energies are expressed as the masses of  these quasi-particles. Figure~\ref{fig:excitations} illustrates low-energy excitations around $q\,{=}\,0$. Because of the transverse staggered field $h$ induced by the external magnetic field, the gapless excitations at $q\,{=}\,0$, $\pi$, ${\pm}\,q_0$ and $\pi {\pm}\,q_0$ for $h\,{=}\,0$ have finite gaps.

The soliton mass $M_{\rm s}$ was calculated analytically by Essler {\it et al.}\,~\cite{Essler1} as
\begin{eqnarray}
\frac{M_{\rm s}}{J}\,{=}\,\frac{2v}{\sqrt{\pi}}\frac{{\Gamma}{\left(\displaystyle\frac{\xi}{2}\right)}}{{\Gamma}{\left(\displaystyle\frac{1\,{+}\,\xi}{2}\right)}}{\left[\frac{{\Gamma}{\left(\displaystyle\frac{1}{1\,{+}\,\xi}\right)}}{{\Gamma}{\left(\displaystyle\frac{\xi}{1\,{+}\,\xi}\right)}} \frac{c{\pi}g{\mu}_{\rm B}H}{2vJ}c_{\rm s}\right]^{(1\,{+}\,\xi)/2}},\ 
\label{eq:solitonmass}
\end{eqnarray}
where $v$ is the dimensionless spin velocity, $\xi$ is a parameter given by ${\xi}\,{=}\,[2/({\pi}R^2)-1]^{-1}$ and $c$ is a parameter depending on the magnetic field. The field dependences of these parameters are given in the literature;\cite{Affleck,Essler1,Hikihara} $v\,{\rightarrow}\,{\pi}/2$, $\xi\,{\rightarrow}\,1/3$ and $c\,{\rightarrow}\,1/2$ for $H\,{\rightarrow}\,0$. This result is applicable in a wide magnetic field range up to the saturation field given by $H_{\rm s}\,{=}\,2J/g{\mu}_{\rm B}$. 

The breathers $M_n$ $(n\,{=}\,1,\,2,\,\cdots)$ corresponding to the excitations at $q\,{=}\,0$ and ${\pi}$ have hierarchical structures labeled by integer $n$. The mass of the $n$th breather is given in terms of the soliton mass $M_{\rm s}$ and parameter $\xi$ as
\begin{eqnarray}
M_n=2M_{\rm s} {\sin}\,\left(\frac{n{\pi}{\xi}}{2}\right).
\label{eq:breather}
\end{eqnarray}
The number of breathers is limited by $n\,{\leq}\,[{\xi}^{-1}]$.\cite{Affleck} In a low magnetic field of $g{\mu}_{\rm B}H/J\,{<}\,0.5$, breathers up to the third order can exist. The soliton resonance labeled as $E_{\rm s}$ in Fig.\,\ref{fig:excitations} corresponds to the $q\,{=}\,0$ excitation on the excitation branch connected to the soliton and antisoliton. Its energy is given by $E_{\rm s}\,{\simeq}\,[{M_{\rm s}}^2+{(g\mu_{\rm B}H)}^2]^{1/2}$.\cite{Affleck} 

The specific heat in SG field theory can be
obtained exactly by taking advantage of its integrability. 
It has been evaluated by two approaches: the thermodynamic Bethe ansatz (TBA)\,\cite{TakahashiSuzuki,Fowler}
and quantum transfer matrix (QTM) methods.\cite{Destri,Essler2} In the TBA method, a set of integral equations is derived from the Bethe ansatz. The number of integral equations is finite
only at special values of the compactification radius,
including the SU(2) symmetric point $R=1/\sqrt{2\pi}$.
Thus, the application to the present problem is practically
limited to the SU(2) symmetric radius, where
the first breather mass $M_1$ degenerates with soliton mass $M_s$.
Precisely speaking, however, the uniform field $H$ breaks the
SU(2) symmetry and thus renormalizes
the compactification radius $R$.
This renormalization effect can be taken into account by
using the QTM method, which allows the calculation of specific
heat for an arbitrary radius. These calculations of the exact specific heat in SG
field theory has been successfully applied the analysis in 
the experimental data on
PM$\cdot$Cu(NO$_3$)$_2\cdot$(H$_2$O)$_2$\,\cite{Feyerherm,Essler2}
and Yb$_4$As$_3$,\cite{Oshikawa2} which exhibit a field-induced
gap similar to that in the present system.

\section{Experimental Details}

KCuGaF$_6$ single crystals were grown by the horizontal Bridgman method from the melt of an equimolar mixture of KF, CuF$_2$ and GaF$_3$ packed into a Pt tube of 10 or 15 mm inner diameter and 100 mm length. The materials were dehydrated by heating in vacuum at about 100$^{\circ}$C for three days. One end of the Pt tube was welded and the other end was tightly folded with pliers. The temperature at the center of a furnace was set at 750$^{\circ}$C, and was decreased at a rate of 1$^{\circ}$C/h to 500$^{\circ}$C. Transparent light-pink crystals with a maximum size of $10\,{\times}\,15\,{\times}\,5$ mm$^3$ were obtained. These crystals were identified as KCuGaF$_6$ by X-ray powder diffraction analysis. 

The crystallographic $a$, $b$ and $c$ axes were determined by X-ray single-crystal diffraction. KCuGaF$_6$ crystals are cleaved along the $(1,\,1,\,0)$ plane. The magnetic susceptibilities ${\chi}_a, {\chi}_b$ and ${\chi}_c$ are anisotropic below 50 K because of the DM interaction.\cite{Umegaki} The magnitudes of the susceptibilities below 50 K decrease in the order ${\chi}_c\,{>}\,{\chi}_b\,{>}\,{\chi}_a$. Thus, the crystallographic axes can be identified from the susceptibility measurements. Specific heat measurements were carried out down to 0.35 K in a magnetic field of up to 9 T using a Physical Property Measurement System (Quantum Design, PPMS). Magnetic fields were applied along the $a, b$ and $c$ axes.


\section{Results and discussion}
\begin{figure}[htbp]
\begin{center}
 \includegraphics[scale =0.45]{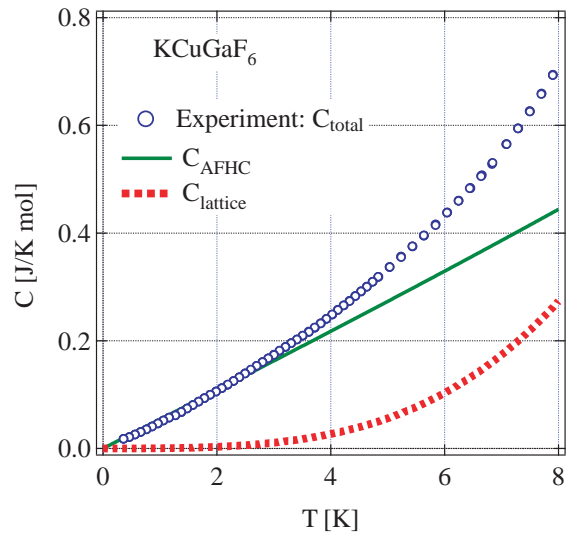}
\end{center}
\caption{Low-temperature total specific heat $C_{\rm total}$ measured at zero magnetic field. Specific heat data are plotted as open circles. $C_{\rm AFHC}$ is the theoretical low-temperature specific heat given by eq.\,(\ref{eq:C_AFHC}) with $J/k_{\rm B}\,{=}\,103$ K, and $C_{\rm lattice}$ is the lattice contribution estimated by subtracting $C_{\rm AFHC}$ from $C_{\rm total}$.} 
\label{fig:sh_AFHC}
\end{figure}

Figure~\ref{fig:sh_AFHC} shows the low-temperature total specific heat $C_{\rm total}$ measured at zero magnetic field. No magnetic ordering was observed down to 0.35 K, which verifies the good one-dimensionality of the present system. $C_{\rm total}$ is almost proportional to temperature below 4 K, which is a characteristic of the $S\,{=}\,1/2$ AFHC.\cite{Kluemper,Johnston} To evaluate the lattice contribution, we used the theoretical specific heat of the $S\,{=}\,1/2$ AFHC.\cite{Kluemper, Johnston} The solid line in Fig.\,\ref{fig:sh_AFHC} indicates the theoretical specific heat $C_{\rm AFHC}$ calculated with $J/k_{\rm B}\,{=}\,103$ K.\cite{Morisaki,Umegaki} For $k_{\rm B}T/J\,{<}\,0.1$, the specific heat of the $S\,{=}\,1/2$ AFHC is approximately expressed as
\begin{eqnarray}
C_{\rm AFHC} = \frac{2Rk_{\rm B}T}{3J},
\label{eq:C_AFHC}
\end{eqnarray}
where $R$ is the gas constant and the spin state is equivalent to the Tomonaga-Luttinger liquid (TLL). $C_{\rm total}$ is the total of the magnetic $C_{\rm mag}$ and lattice $C_{\rm lattice}$ contributions. The lattice contribution $C_{\rm lattice}$ shown by the dashed line in Fig.\,\ref{fig:sh_AFHC} was obtained by subtracting $C_{\rm AFHC}$ from $C_{\rm total}$. The magnetic specific heat in a finite magnetic field was obtained by subtracting $C_{\rm lattice}$ from the total specific heat $C_{\rm total}$.

\begin{figure}[htbp]
\begin{center}
\includegraphics[scale =0.45]{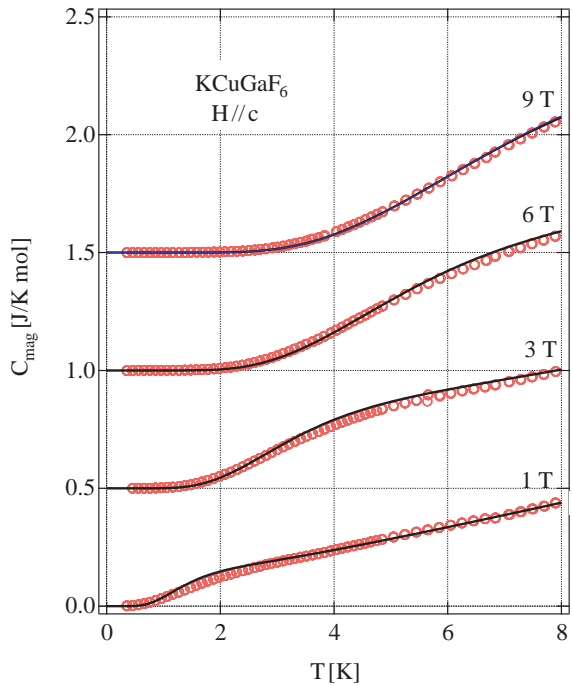}
\end{center}
\caption{Temperature dependence of magnetic specific heat $C_{\rm mag}$ measured at $H\,{=}\,1, 3, 6$ and 9 T for $H\,{\parallel}\,c$. Each set of data is shifted upward by a multiple of 0.5 J/(K mol). Open circles indicate experimental data and solid lines are the theoretical specific heat $C_{\rm TBA}({\Delta})$ based on quantum SG field theory fitted with the soliton mass ${\Delta}$ shown in Fig.\,\ref{fig:soliton_gap_c}.} 
\label{fig:sh_c}
\end{figure}

Figure~\ref{fig:sh_c} shows the low-temperature magnetic specific heat $C_{\rm mag}$ obtained at various magnetic fields parallel to the $c$ axis. For this field direction, the proportionality coefficient $c_{\rm s}\,{=}\,h/H$ is the largest among the three directions and $c_{\rm s}\,{=}\,0.17$. In a finite magnetic field, $C_{\rm mag}$ increases exponentially with increasing temperature, which indicates the emergence of a magnetic-field-induced gap. With further increasing temperature, $C_{\rm mag}$ exhibits a rounded shoulder and increases linearly. The shoulder shifts to a higher temperature and becomes broader as the magnetic field increases. This shows that the gap increases with the magnetic field. 

We analyzed the experimental results using the exact 
specific heat in SG field theory.
To verify the accuracy of the calculation and to determine the effect of
the field renormalization of $R$, we compared the experimental
data with the theoretical specific heats obtained by the TBA and QTM methods.
In the TBA calculation, $R$ is fixed to the SU(2) symmetric radius
as discussed above. In the QTM calculation, $R$ is obtained as
a function of $H$ using the exact Bethe ansatz solution
of the Heisenberg chain without the staggered field.\cite{KIB}
Thus, the only adjustable fitting parameter is the soliton mass
$M_{\rm s}$ in both calculations.

\begin{figure}[htbp]
\begin{center}
\includegraphics[scale =0.45]{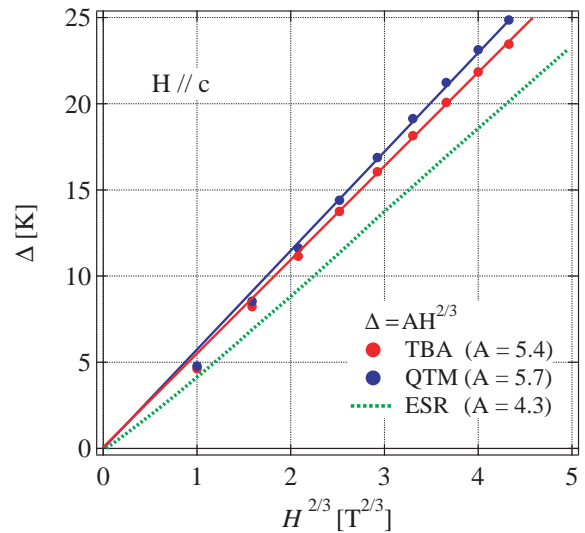}
\end{center}
\caption{Soliton mass ${\Delta}\,({\equiv}\,M_{\rm s})$ for $H\,{\parallel}\,c$ as a function of $H^{2/3}$ obtained from the fits of the theoretical specific heat calculated using the TBA and QTM methods. The dotted line shows the soliton mass obtained from the ESR measurements.\cite{Umegaki}} 
\label{fig:soliton_gap_c}
\end{figure}

The solid lines in Fig.~\ref{fig:sh_c} show the fits
by $C_{\rm TBA}({\Delta})$ with the soliton mass shown in
Fig.~\ref{fig:soliton_gap_c}. On the whole, the experimental specific heat
for $H\,{\parallel}\,c$ is well described by the theory based on
quantum SG field theory, although some discrepancy is
observed. The following three factors may have led to the discrepancy
between the experimental and theoretical results: the breaking of the
SU(2) symmetry, the error in the estimation of the lattice contribution
$C_{\rm lattice}$ and the effect of unknown excitation modes observed in previous ESR measurements.\cite{Umegaki}

\begin{figure}[htbp]
\begin{center}
\includegraphics[scale =0.45]{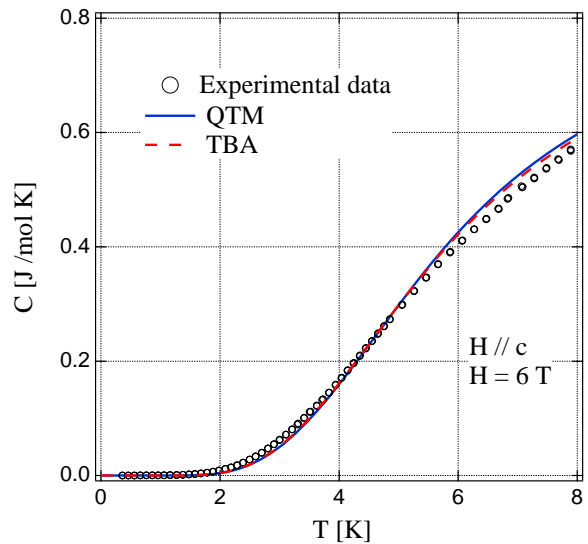}
\end{center}
\caption{Analyses of the specific heat obtained using the TBA and QTM methods based on quantum SG field theory. Circles show the magnetic specific heat measured at $H\,{=}\,6$\,T for $H\,{\parallel}\,c$. Dashed and solid lines indicate the theoretical specific heat calculated using the TBA and QTM methods, respectively.} 
\label{fig:sh_theory}
\end{figure}

The effect of breaking the SU(2)
symmetry is determined by comparing the TBA calculation
with $R$ fixed at $1/\sqrt{2\pi}$ and the QTM calculation with the renormalized $R$.
Figure~\ref{fig:sh_theory} shows a comparison between the results of 
analyses based on the TBA and QTM methods. As shown in
Figs.\,\ref{fig:soliton_gap_c} and \ref{fig:sh_theory}, the temperature
dependences of the specific heat and the soliton masses obtained from
both methods are similar.
This means that, in the present magnetic field range,
the effect of the renormalization of $R$ is not important and
the system can be well approximated by the SU(2) symmetric
radius $R=1/\sqrt{2\pi}$.
In fact, in the present material,
the exchange interaction $J/k_{\rm B}\,{=}\,103$ K
is dominant over the magnetic field
used in the experiments.\cite{Morisaki,Umegaki}
In addition, the specific heat is not very sensitive to
a small change in $R$.

The soliton masses $\Delta$ for $H\,{\parallel}\,c$ evaluated from the analyses using the TBA and QTM methods are shown in Fig.~\ref{fig:soliton_gap_c} as a function of $H^{2/3}$. It is clear that $\Delta$ is proportional to $H^{2/3}$. This relation can be derived by setting ${\xi}\,{=}\,1/3$ in eq.\,(\ref{eq:solitonmass}). Note that in the linear spin wave theory, the energy gap exhibits the field dependence of ${\Delta}\,{\propto}\,H^{1/2}$. Therefore, the field dependence of the soliton mass for $H\,{\parallel}\,c$ agrees well with that in quantum SG field theory. 
  
The dotted line in Fig.~\ref{fig:soliton_gap_c} indicates the soliton mass $\Delta$ obtained from previous ESR measurements.\cite{Umegaki} The soliton mass obtained from the specific heat measurements is 1.26 times as large as that obtained from the ESR measurements. The overestimation of the soliton mass is attributed to the unknown higher-energy excitation modes $U_2$ and $U_3$ observed in ESR measurements.\cite{Umegaki} In another quantum SG system PM$\cdot$Cu(NO$_3$)$_2\cdot$(H$_2$O)$_2$, the ratio of the gap obtained from specific heat measurements\,\cite{Feyerherm} to that obtained from ESR measurements\,\cite{Zvyagin} is similar to that observed for KCuGaF$_6$. While the small disagreement in the soliton mass is found, the experimental $C_{\rm mag}$ for $H\,{\parallel}\,c$ is well described within the framework of quantum SG field theory, as shown in Figs.~\ref{fig:sh_c} and \ref{fig:sh_theory}.

\begin{figure}[htbp]
\begin{center}
 \includegraphics[scale =0.45]{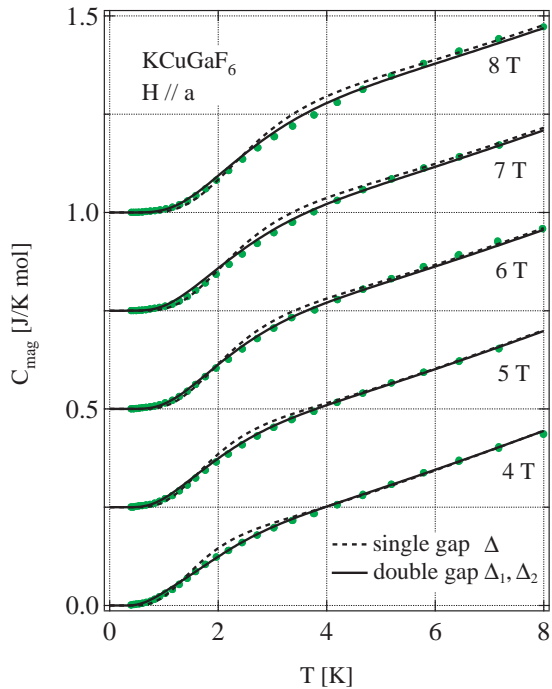}
\end{center}
\caption{Magnetic specific heat $C_{\rm mag}$ measured for $H\,{\parallel}\,a$. Each set of data is shifted upward by a multiple of 0.25 J/(K mol). Dashed lines denote fits by $C_{\rm TBA}({\Delta})$ with a single soliton mass $\Delta$, while solid lines indicate fits using eq.~(\ref{eq:heat2}) with the two gaps $\Delta_1$ and $\Delta_2$ shown in Fig.~\ref{fig:D12}. }
\label{fig:H_para_a}
\end{figure} 

\begin{figure}[htbp]
\begin{center}
 \includegraphics[scale =0.45]{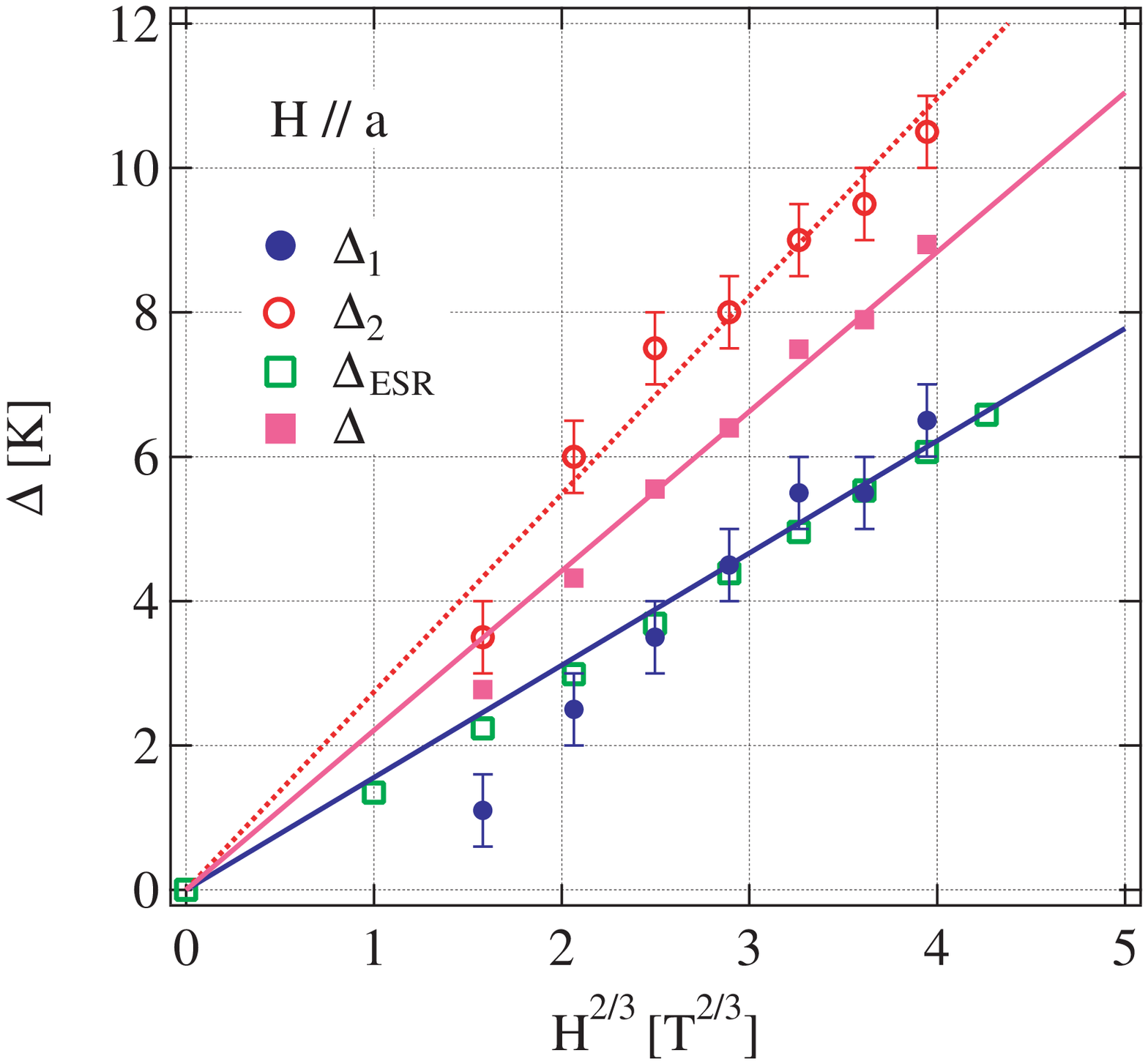}
\end{center}
\caption{Soliton mass ${\Delta}\,({\equiv}\,M_{\rm s})$ for $H\,{\parallel}\,a$ as a function of $H^{2/3}$ obtained from fits of the theoretical specific heat $C_{\rm TBA}({\Delta})$ (closed squares). Closed and open circles indicate the gaps ${\Delta_1}$ and $\Delta_2$ obtained by fits using eq.~(\ref{eq:heat2}), respextively. Open squares indicate the soliton mass estimated by ESR measurements. }
\label{fig:D12}
\end{figure}

Next, we show in Fig.~\ref{fig:H_para_a} the low-temperature magnetic specific heat $C_{\rm mag}$ obtained at various magnetic fields for $H\,{\parallel}\,a$. For this field direction, the proportionality coefficient is the smallest and $c_{\rm s}\,{=}\,0.031$.\cite{Umegaki} From the exponential increase in $C_{\rm mag}$, we see that the gap opens in a magnetic field. However, because of the small proportionality coefficient $c_{\rm s}$, the magnitude of the gap is much smaller than that for $H\,{\parallel}\,c$ at the same magnetic field. First, we fitted the theoretical specific heat $C_{\rm TBA}({\Delta})$ to the experimental data. The soliton mass $\Delta$ obtained is plotted in Fig.~\ref{fig:D12} as a function of $H^{2/3}$. As shown by the dashed lines in Fig.~\ref{fig:H_para_a}, some discrepancy was observed between the experimental and theoretical results. A similar discrepancy was also observed in copper benzoate when an external field was applied along the $a^{\prime\prime}$ direction, for which the gap becomes smallest.\cite{Essler2} Although $\Delta$ is proportional to $H^{2/3}$, as predicted by quantum field theory, its magnitude is 1.45 times as large as that obtained from the ESR measurements. It is considered that this discrepancy for KCuGaF$_6$ arises from the higher-energy unknown mode $U_1$ and multiple excitation modes $C_n$, which cannot be explained within the framework of quantum SG field theory.\cite{Umegaki} 

The energy of the $U_1$ mode observed for $H\,{\parallel}\,a$ is larger than the mass of the first breather and is comparable to those of the second and third breathers. The $C_n$ modes are multiple excitations of the $E_{\rm s}$ and the breathers, whose energies are given by $E_{\rm s}\,{+}\,M_n$. Note that these modes were also observed in ESR measurements on PM$\cdot$Cu(NO$_3$)$_2\cdot$(H$_2$O)$_2$.\cite{Zvyagin} These excitations can contribute to the magnetic specific heat because they are excitations from the ground state. Then, assuming that the contribution of these higher-energy excitations is effectively represented as $C_{\rm TBA}(\Delta_2)$ with the secondary gap $\Delta_2$, we express the theoretical specific heat as 
\begin{eqnarray}
C_{\rm mag}^{\rm eff}=\frac{1}{2}\left[C_{\rm TBA}(\Delta_1)+C_{\rm TBA}(\Delta_2)\right].
\label{eq:heat2}
\end{eqnarray}
The first term expresses the contribution of the excitations of the quantum SG model, while the second term expresses the contribution of other higher-energy excitations. The primary gap $\Delta_1$ corresponds to the true soliton mass. The solid lines in Fig.~\ref{fig:H_para_a} show fits using eq.\,(\ref{eq:heat2}). The experimental specific heat in various magnetic fields for $H\,{\parallel}\,a$ ia well described by $C_{\rm mag}^{\rm eff}$. The two gaps, $\Delta_1$ and $\Delta_2$, obtained from the fits are plotted as functions of $H^{2/3}$ in Fig.~\ref{fig:D12} together with the soliton mass estimated from the ESR measurements.\cite{Umegaki} Both energy gaps are proportional to $H^{2/3}$ for $H\,{\geq}\,3$ T. The primary gap $\Delta_1$ coincides with the soliton mass estimated from the previous ESR measurements. This means that the lower excitations for $H\,{\parallel}\,a$ can be understood within the framework of quantum SG field theory. 

\begin{figure}[htbp]
\begin{center}
\includegraphics[scale =0.45]{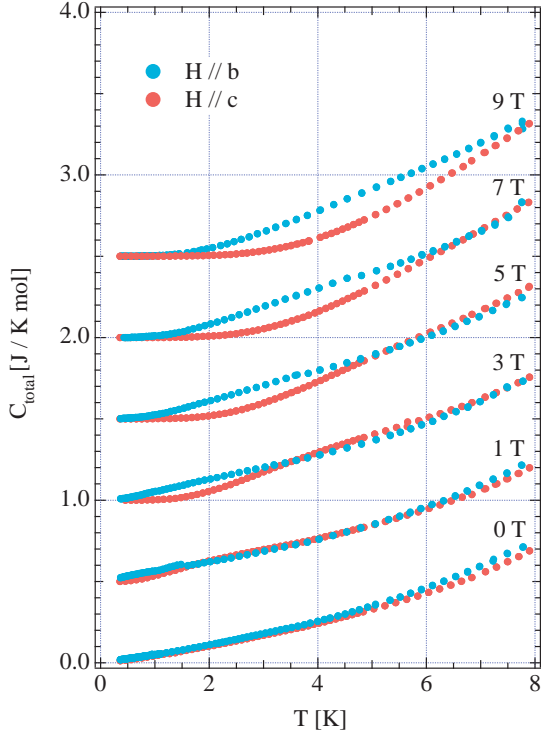}
\end{center}
\caption{Comparison between total specific heat $C_{\rm total}$ for $H\,{\parallel}\,b$ and $H\,{\parallel}\,c$ measured at various magnetic fields. Each set of data is shifted upward by a multiple of 0.5 J/(mol K).} 
\label{fig:sh_b_c}
\end{figure}

\begin{figure}[htbp]
\begin{center}
\includegraphics[scale =0.45]{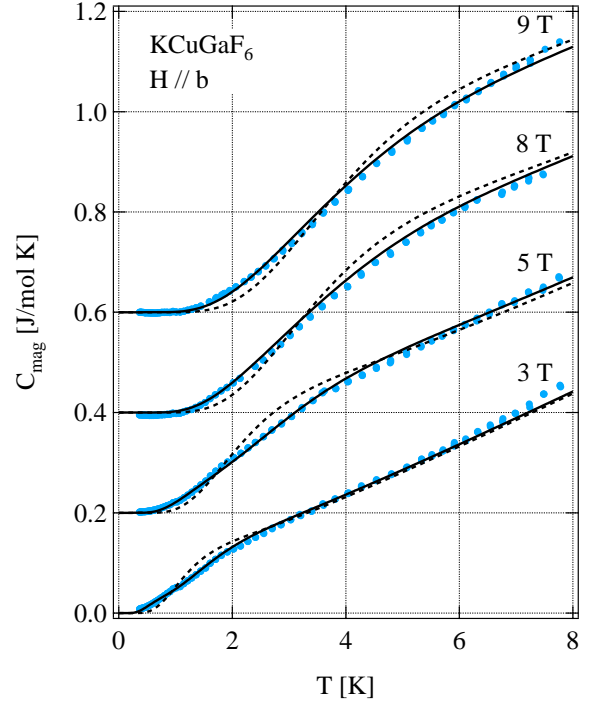}
\end{center}
\caption{Magnetic specific heat $C_{\rm mag}$ measured for $H\,{\parallel}\,b$. Each set of data is shifted upward by a multiple of 0.2 J/(mol K). Dashed lines show fits by the theoretical specific heat $C_{\rm TBA}({\Delta})$, while solid lines indicate fits using eq.~(\ref{eq:heat2}).} 
\label{fig:sh_b}
\end{figure}

Figure~\ref{fig:sh_b_c} shows the temperature dependences of the total specific heat $C_{\rm total}$ for $H\,{\parallel}\,b$ and $H\,{\parallel}\,c$, for which the proportionality coefficients are $c_{\rm s}\,{=}\ $0.16 and 0.17, respectively.\cite{Umegaki} The values of magnetic specific heat for these two field directions were expected to be similar, because the proportionality coefficients are close to each other. However, the values of specific heats are clearly different, as shown in Fig. \ref{fig:sh_b_c}. It is apparent that the magnetic-field-induced gap for $H\,{\parallel}\,b$ is much smaller than that for $H\,{\parallel}\,c$. A notable feature is that the specific heat for $H\,{\parallel}\,b$ is almost linear in temperature for $H\,{\leq}\,2$\,T. This indicates that the ground state is gapless at low magnetic fields. The low-magnetic-field specific heat dose not exhibits an anomaly indicative of 3D ordering down to 0.35\,K, although the ground state appears gapless. These observations appear to be consistent with the result of previous ESR measurements for $H\,{\parallel}\,b$, where we observed an intense unknown $U_4$ mode with resonance frequency proportional to $H\,{-}\,H_{\rm c}$ with $H_{\rm c}\,{\simeq}\,2.5$\,T.\cite{Umegaki}

\begin{figure}[htbp]
\begin{center}
 \includegraphics[scale =0.45]{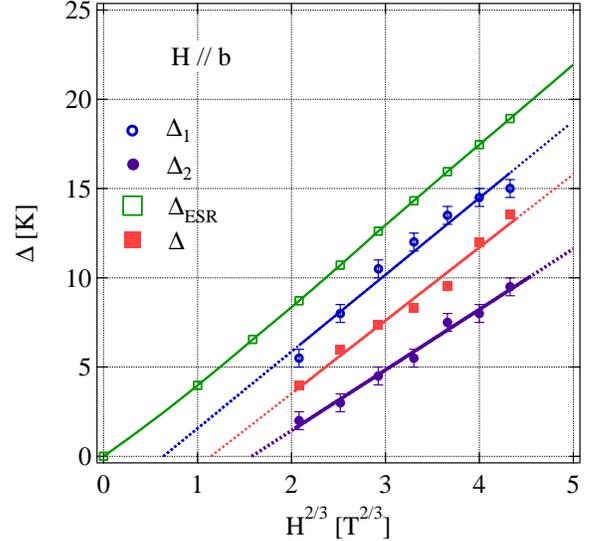}
\end{center}
\caption{Soliton mass ${\Delta}\,({\equiv}\,M_{\rm s})$ for $H\,{\parallel}\,b$ as a function of $H^{2/3}$ obtained from the fits of the theoretical specific heat $C_{\rm TBA}({\Delta})$ (closed squares). Closed and open circles indicate the gaps ${\Delta_1}$ and $\Delta_2$ obtained by fits using eq.~(\ref{eq:heat2}), respectively. Open squares indicate the soliton mass obtained from ESR measurements. }
\label{fig:D12_b}
\end{figure}  

\begin{figure}[htbp]
\begin{center}
 \includegraphics[scale =0.45]{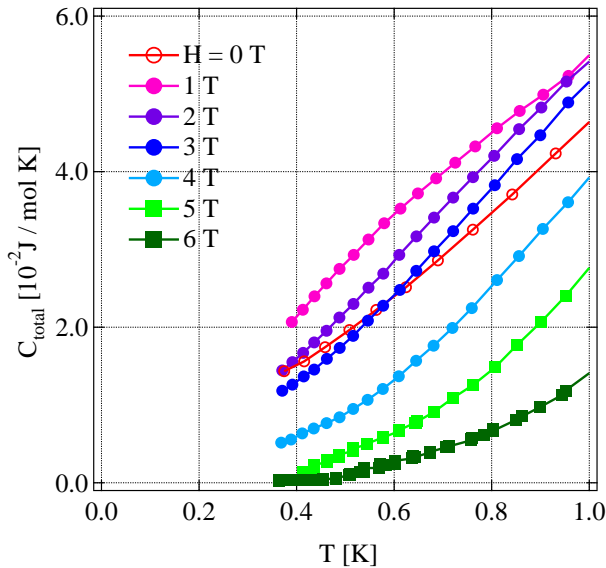}
\end{center}
\caption{Low-temperature total specific heat $C_{\rm total}$ measured at various magnetic fields for $H\,{\parallel}\, b$. Open circles indicate $C_{\rm total}$ at zero field. Closed symbols indicate $C_{\rm total}$ in nonzero magnetic fields.
}
\label{fig:Cmag_lowT_b}
\end{figure} 

\begin{figure}[htbp]
\begin{center}
 \includegraphics[scale =0.45]{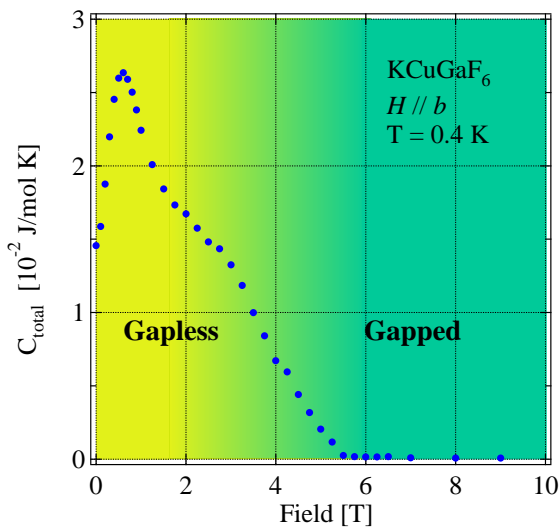}
\end{center}
\caption{Magnetic field dependence of total specific heat $C_{\rm total}$ measured at $T\,{=}\,0.4$ K for $H\,{\parallel}\,b$. The quantum phase transition where a gapless ground state changes into a gapped state is observed.}
\label{fig:C_H_b}
\end{figure} 
  
Figure~\ref{fig:sh_b} shows the magnetic specific heat measured at various magnetic fields for $H\,{\parallel}\,b$. Fitting of the theoretical specific heat $C_{\rm TBA}({\Delta})$ to the experimental data is not successful, as shown by the dashed lines in Fig.~\ref{fig:sh_b}, and contrary to the case of $H\,{\parallel}\,a$, we obtain a soliton mass $\Delta$ considerably smaller than that estimated from ESR measurements (see Fig.~\ref{fig:D12_b}). Soliton mass $\Delta$ is not proportional to $H^{2/3}$ but can be expressed as ${\Delta}\,{=}\,A\{H^{2/3}\,{-}\,H_{\rm c}(0)^{2/3}\}$ with $H_{\rm c}(0)\,{=}\,1.22$\,T. In previous ESR measurements, we observed three unknown modes, $U_1, U_2$ and $U_3$, whose excitation energies are close to or lower than the first breather mass added to in the $U_4$ mode.\cite{Umegaki} The unexpectedly small soliton mass obtained from the present specific heat measurements should be due to these low-energy unknown modes. Then, assuming that the contribution of these unknown modes is effectively expressed in terms of the secondary gap $\Delta_2$, which is much smaller than the soliton mass $\Delta_1$, we describe the magnetic specific heat by eq.~(\ref{eq:heat2}), as in the case of $H\,{\parallel}\,a$. The solid lines in Fig.~\ref{fig:sh_b} indicate the fits using eq.~(\ref{eq:heat2}) with the two gaps shown in Fig.~\ref{fig:D12_b}. the experimental value of $C_{\rm mag}$ for $H\,{\parallel}\,b$ is effectively described by eq.~(\ref{eq:heat2}). Similar to the soliton gap $\Delta$ obtained from the fit by $C_{\rm TBA}({\Delta})$, both the primary gap $\Delta_1$ and the secondary gap $\Delta_2$ obtained for $H\,{\geq}\,3$ T are expressed as ${\Delta}_i\,{=}\,A\{H^{2/3}\,{-}\,H_{\rm c}(i)^{2/3}\}$ $(i\,{=}\,1$ and 2) with $H_{\rm c}(1)\,{=}\,0.51$\,T and $H_{\rm c}(2)\,{=}\,1.98$\,T. This value of $H_{\rm c}(2)$ is close to the critical field $H_{\rm c}\,{\simeq}\,2.5$\,T observed in ESR measurements for $H\,{\parallel}\,b$.\cite{Umegaki} 

In Fig.~\ref{fig:Cmag_lowT_b}, we show the total specific heat $C_{\rm total}$ measured below 1 K at various magnetic fields. In this temperature range, the lattice contribution $C_{\rm lattice}$ is negligible. $C_{\rm total}$ appears to be linear in temperature for $H\,{\leq}\,3$ T, while for $H\,{\geq}\,5$ T, $C_{\rm total}$ exhibits exponential temperature dependence. This indicates that the gapless ground state changes into a gapped state between 3 and 5 T. To confirm the magnetic-field-induced quantum phase transition, we performed a field scan of specific heat at $T\,{=}\,0.4$\,K. The result is shown in Fig.~\ref{fig:C_H_b}. The total specific heat $C_{\rm total}$ has a peak at $H\,{=}\,0.6$ T and a shoulder at $H\,{=}\,3$ T. This behavior was not observed for $H\, {\parallel}\,a$ and $H\, {\parallel}\,c$. With further increasing magnetic field, $C_{\rm total}$ decreases rapidly and becomes almost zero at $H\,{=}\,5.5$ T. From these observations, we can deduce that the gapless ground state  for $H\,{\parallel}\,b$ becomes a gapped state at a critical field $H_{\rm c}$ of 3 and 5.5 T. The critical field is larger than $H_{\rm c}\,{\simeq}\,2.5$\,T observed in ESR measurements.\cite{Umegaki} At present, the origins of the low-field gapless state for $H\,{\parallel}\,b$ and the unknown ESR modes that make a large contribution to the specific heat are not clear. The uniform $b$ axis component of the $\bm D$ vector, which gives rise to a helical spin structure or soliton lattice in a magnetic field,\cite{Garate} may be responsible for these effects.

\section{Conclusion}
We have presented results for the specific heat of KCuGaF$_6$ in magnetic fields. It was clearly observed that an excitation gap opens in a magnetic field and increases with increasing magnetic field. Specific heat data were analyzed using two theoretical values of specific heat calculated by the TBA and QTM methods, both of which are based on the quantum SG model. In the TBA method, the SU(2) symmetry is assumed, while in the QTM method, the breaking of the SU(2) symmetry owing to the external magnetic field is taken into account. The specific heat for $H\,{\parallel}\,c$ is well reproduced by these two calculations with the soliton mass close to that obtained from previous ESR measurements. The soliton mass was found to be proportional to $H^{2/3}$. These results indicate that the thermodynamic properties for $H\,{\parallel}\,c$ are well described by quantum SG field theory, and that the breaking of the SU(2) symmetry is negligible in our experimental magnetic field range because of the large exchange interaction of $J/k_{\rm B}\,{=}\,103$\,K.

In the cases of $H\,{\parallel}\,a$ and $H\,{\parallel}\,b$, we found a significant contribution of additional excitations that cannot explained in the framework of quantum SG field theory. For these field directions, we analyzed the specific heat data using eq.~(\ref{eq:heat2}), in which the contribution of the unknown excitations is effectively taken into account by assuming a secondary gap ${\Delta}_2$. For $H\,{\parallel}\,a$, the primary gap ${\Delta}_1$ agrees with the soliton mass ${\Delta}_{\rm ESR}$ obtained from the ESR measurements, while for $H\,{\parallel}\,b$, ${\Delta}_1$ does not agree with ${\Delta}_{\rm ESR}$. We found that for $H\,{\parallel}\,b$, the ground state is gapless at low magnetic fields, and that with increasing magnetic field, a quantum phase transition occurs between gapless and gapped ground states.


\begin{acknowledgments}
The authors would like to acknowledge H. Nojiri, S. C. Furuya and I. Affleck for useful discussions. This work was supported by the Grant-in-Aid for Scientific Research (A) from the Japan Society for the Promotion of Science, and by a Global COE Program ``Nanoscience and Quantum Physics'' at Tokyo Tech., both funded by the Japanese Ministry of Education, Culture, Sports, Science and Technology. I.U. and H.T. were supported by a JSPS Research Fellowship for Young Scientists and a grant from the Mitsubishi Foundation, respectively.
\end{acknowledgments}


\end{document}